\documentstyle[12pt]{article}
\newcommand{\beq}{\begin{equation}}
\newcommand{\eeq}{\end{equation}}
\newcommand{\beqn}{\begin{eqnarray}}
\newcommand{\eeqn}{\end{eqnarray}}
\newcommand{\slp}{\raise.15ex\hbox{$/$}\kern-.57em\hbox{$\partial$}}
\newcommand{\slA}{\raise.15ex\hbox{$/$}\kern-.57em\hbox{$A$}}
\newcommand{\lnA}{\raise.15ex\hbox{$/$}\kern-.57em\hbox{$A$}}
\newcommand{\bP}{\bar{\Psi}}
\newcommand{\phiH}{\hat{\phi}}
\newcommand{\etaH}{\hat{\eta}}
\newcommand{\hs}{\hspace*{0.6cm}}

\begin{document}

\title{Exact electronic Green functions in a Luttinger liquid with long-range interactions}
\author{
An\'{\i}bal Iucci$^{a}$, Carlos Na\'on$^{a}$}
\date{December 1999}
\maketitle

\def\thepage{\protect\raisebox{0ex}{\ } La Plata-Th 99/15}
\thispagestyle{headings}
\markright{\thepage}

\begin{abstract}

\hs We compute the 2-point (equal-time) electronic Green function
in a Tomonaga-Luttinger system with long range electron-electron
interactions. We obtain an analytical expression for a "super
long-range" potential of the form
$V(x)=\frac{e^2d^{-\epsilon}}{|x|^{1-\epsilon}}$. As a consistency
check of our computational technique we also consider the
particular case of a Coulomb potential. Our result confirms the
$\exp-C(logx)^\frac{3}{2}$ long-distance behavior first obtained
by Schulz.
\end{abstract}

\vspace{3cm} Pacs: 05.30.Fk\\
\hspace*{1,7cm}71.10.Pm

\noindent --------------------------------

\noindent $^a$ {\footnotesize Departamento de F\'{\i}sica.
Universidad Nacional de La Plata.  CC 67, 1900 La Plata,
Argentina.\\ e-mails: iucci@venus.fisica.unlp.edu.ar;
naon@venus.fisica.unlp.edu.ar}

\newpage
\pagenumbering{arabic}

\hs Models of one-dimensional (1d) interacting electrons are
relevant as testing grounds for new ideas and can help us to
understand systems in higher dimensions. In particular they are
useful to describe the behavior of strongly anisotropic physical
systems in condensed matter, such as organic conductors
\cite{organic conductors}, charge transfer salts \cite{salts} and
quantum wires \cite{quantum wires}. Probably the two most widely
studied 1d systems are the Hubbard model \cite{Hubbard} and the
so-called "g-ology" model \cite{Solyom}. They are known to display
the Luttinger liquid \cite{Haldane} behavior characterized by
spin-charge separation and by non-universal (interaction
dependent) power-law correlation functions. In these models one
usually considers short-range electron-electron interactions. This
picture works well for conductors in which the screening between
adjacent chains reduces the range of interactions within one chain
\cite{Schulz1}. On the other hand, as the dimensionality of the
system decreases charge screening effects are expected to become
less important and the long-range interaction between electrons
seems to play a central role in determining the properties of the
system. This assertion seems to be confirmed by experiments in
GaAs quantum wires \cite{quantum wires} and quasi-1d conductors
\cite{quasi-exp}. From a theoretical point of view the effects of
long-range interactions has been also recently discussed in
connection to a variety of problems such as the Fermi-edge
singularity \cite{Fermi-edge}, the insulator-metal transition
\cite{insulator-metal} and the role of the lattice through umklapp
scattering and size dependent effects \cite{lattice}.

The effect of Coulomb forces on the single-particle Green function
and on charge-density correlations in 1d systems have been
previously investigated in a pioneering work by Schulz
\cite{Schulz2} using the conventional bosonization method
\cite{conventional bosonization}. The purpose of the present paper
is to compute the electronic Green function in the presence of
long-range interactions using an alternative path-integral
bosonization technique previously developed in the context of
quantum field theories \cite{path-integral bosonization}. We
consider a non-local version of the Thirring model \cite{non local
Thirring} described by the following (Euclidean) action

\beqn S=\int d^2x \bP i\slp\Psi- \int d^2x d^2y \left[V_{(0)}(x,y)
J_0(x)J_0(y) + V_{(1)}(x,y) J_1(x) J_1(y) \right],
\label{eq:ThirringAction} \eeqn

\noindent where $x= ({\bf x},\tau_{x})$, and
$J_0={\bP}\gamma_{0}{\Psi}=\rho$ and $J_1={\bP}\gamma_{1}{\Psi}=j$
are the charge density and current density operators respectively.
Let us also mention that, for simplicity, we shall take $v_F=1$
and $p_F=0$ throughout this article. The functions
$V_{(\mu)}(x,y)$ are forward-scattering potentials. Setting
$V_{(0)}=V_{(1)}=\frac{g^2}{2}\delta^2(x-y)$ one gets the
covariant and local version of the Thirring model usually studied
in the context of (1+1) QFT's. On the other hand, the choice \beqn
V_{(0)}(x,y)&=&V(|{\bf x}-{\bf y}|) \,\
\delta(\tau_x-\tau_y),\nonumber\\ V_{(1)}(x,y)&=&0,
\label{eq:AnyLR} \eeqn yields the simplest version of the
Tomonaga-Luttinger (TL) model with an instantaneous distance
dependent potential and no current-current fluctuations. The main
purpose of the present work is to study this case for a long-range
electron-electron potential, {\it different from the Coulomb
interaction}, of the form

\beqn V(|{\bf x}-{\bf y}|)&=&\frac{e^2d^{-\epsilon}}{|{\bf x}-{\bf
y}|^{1-\epsilon}} \,\ \delta(\tau_x-\tau_y), \label{eq:SuperLR}
\eeqn

\noindent where $d$ is a length scale and $\epsilon$ is a small
real number ($0<\epsilon\leq1$). This interaction has been
previously analyzed by using renormalization group techniques in
order to explore the possibility of having Luttinger liquid
behavior in more than one dimension \cite{Schmeltzer} and in
connection to the vacuum structure of the non local Thirring model
\cite{Ground}.

\vspace{1cm}

Before considering the "super long range" potential described
above, we shall sketch our functional approach to bosonization in
a general case.
 As shown in ref. \cite{non local Thirring} the
partition function of the model defined by eq.
(\ref{eq:ThirringAction}) can be solved, in the path-integral
framework, by means of a chiral change of fermionic variables.
This procedure allows to obtain a bosonized effective action in
terms of scalar fields that are naturally identified with the
collective modes (plasmons) of the system. Since this method has
been described many times in the literature, here we shall skip
the details (the interested reader is refer to \cite{path-integral
bosonization}, \cite{non local Thirring} and references therein).
Let us only say that using a Hubbard-Stratonovich-like identity,
which amounts to introducing an auxiliary vector field $A_{\mu}$,
the partition function of the general model given by
(\ref{eq:ThirringAction}) can be written in terms of a functional
fermionic determinant as \beqn Z = \int {\cal{D}} A_{\mu}\,\ det(i
\slp + g \lnA )\,\ e^{-S[A]}, \eeqn \noindent with \beqn
S[A]=\frac{1}{2}\int d^2xd^2y
V_{(\mu)}^{-1}(x,y)A_{\mu}(x)A_{\mu}(y), \eeqn \noindent where
$V_{(\mu)}^{-1}$ is such that \beq \int d^2z~ V_{(\mu)}^{-1}(z,x)
V_{(\mu)}(y,z) = \delta^2 (x-y). \eeq Decomposing $A_\mu(x)$ in
longitudinal and transverse pieces
\begin{equation}
A_\mu(x)=\epsilon_{\mu\nu}\partial_\nu\phi(x)+\partial_\mu\eta(x),
\end{equation}

\noindent where $\phi$ and $\eta$ are boson fields (to be
associated to the normal modes of the system) and applying, as
anticipated, functional bosonization techniques to express the
fermionic determinant in terms of $\phi$ and $\eta$, one finally
obtains

\begin{equation}
Z=N \int {\cal{D}} \phi {\cal{D}} \eta \,\ e^{-S_{bos}},
\end{equation}

\noindent where $N$ is a normalization constant and $S_{bos}$,
already written in momentum space, reads
\begin{equation}
S_{bos}=\frac{1}{(2\pi)^2}\int
d^2p\,\left[\phiH(p)\phiH(-p)A(p)+\etaH(p)\etaH(-p)B(p) +
\phiH(p)\etaH(-p)C(p)\right]
\end{equation}

\noindent with
\beq
     A(p) = \frac{1}{\pi}~ p^2 +
     \frac{1}{2}[\hat{V}_{(0)}^{-1}(p) p_1^2 +
           \hat{V}_{(1)}^{-1}(p) p_0^2],
\eeq

\beq B(p) = \frac{1}{2}[\hat{V}_{(0)}^{-1}(p) p_0^2 +
           \hat{V}_{(1)}^{-1}(p) p_1^2],
           \eeq

\beq C(p) = [\hat{V}_{(0)}^{-1}(p) - \hat{V}_{(1)}^{-1}(p)] p_0
p_1, \eeq \noindent and

\beq \Delta = C^2(p)-4A(p)B(p). \eeq

\vspace{1cm}

Let us now focus our attention on the one-particle fermionic
propagator

\beq <\Psi(x) \bP(y)> = \left( \begin{array}{cc}
                        0     &G_R(x,y)\\
                  G_L(x,y) &   0
                  \end{array}   \right).  \\
\eeq

The above depicted bosonization scheme can be easily employed to
evaluate this expression. First of all it is straightforward to
verify that the non-zero components of the Green function
factorize as

\begin{equation}
G_{R,L}(x,y)=G^{(0)}_{R,L}(x,y)B_{R,L}(x,y),
\end{equation}

\noindent where $G^{(0)}_{R,L}(z)=\frac{1}{2\pi(\tau_z^{2}+{\bf
z}^2) }(\tau_z \pm i{\bf z})$ describe free right and left
propagation, and

\beq B_{R,L}(x,y) = \frac{\int D\hat{\Phi} D\hat{\eta}~
e^{-[S_{bos} + S_{R,L}(x,y)]}}{\int D\hat{\Phi} D\hat{\eta}~
e^{-S_{bos}}}. \eeq In this equation one has defined

\begin{equation}
S_{R,L}(x,y)=-\frac{1}{(\pi)^2} \int
d^2p\:[\pm\phiH(p)+i\etaH(p)]\,\ (e^{-ip\cdot x} - e^{-ip\cdot
y}).
\end{equation}

At this point we notice that the evaluation of $B_{R,L}(x,y)$ can
be carried out by means of a convenient shift of the fields
$\phiH(p)$ and $\etaH(p)$. This standard procedure gives

\begin{equation}
B_{R,L}(x,y)=\exp[I_{R,L}(x,y)],
\end{equation}

\noindent where $I_{R,L}$ is a functional of the potentials given
by

\beqn I_{R,L}(x,y)=-\frac{1}{\pi^2}\int
d^2p\sin^2[{\scriptstyle\frac{1}{2}} p\cdot(x-y)] \nonumber\\
\times\frac{\frac{2}{\pi}\hat{V}_{(0)}(p)\hat{V}_{(1)}(p) p^2 +
[\hat{V}_{(0)}(p)-\hat{V}_{(1)}(p)](p_{0}\mp
ip_{1})^2}{\frac{2}{\pi} p^2
[\hat{V}_{(1)}(p)p_{0}^2+\hat{V}_{(0)}(p)p_{1}^2] +
p^4}.\label{I_{R,L}}\eeqn

\noindent This result gives a very general expression for the
2-point electronic correlator as functional of density-density
($\hat{V}_{(0)}(p)$) and current-current ($\hat{V}_{(1)}(p)$)
interaction potentials. It can be used as an alternative route to
analyze the effect of long-range interactions on a Luttinger
system. Of course, if one considers the local and covariant case
$\hat{V}_{(0)}(p)= \hat{V}_{(1)}(p)= \frac{g^2}{2}$ one easily
obtains the well-known Thirring behavior: \beq
B^{Thirring}_{R,L}(x-y) \propto \mid x-y \mid^{-\frac{1}{2}
(g^2/\pi)^2 / (1+ g^2/\pi)}.\label{eq:Thirring} \eeq

Let us now proceed with our main task and specialize formula
(\ref{I_{R,L}}) for the TL model defined by equation
(\ref{eq:AnyLR}). Having $\hat{V}_{(1)}(p)=0$ and
$\hat{V}_{(0)}(p)=\hat{V}_{(0)}(p_{1})$ greatly simplifies the
integrand in (\ref{I_{R,L}}) which allows to perform the integral
in $p_{0}$. Moreover, since we are concerned with the equal-time
correlator, we take $\tau_x=\tau_y$. In this case there is no
distinction between $I_R$ and $I_L$ and one gets
\beqn\label{eq:integral} I_{R,L}({|{\bf x}-{\bf
y}|},\tau_x=\tau_y)=I({|{\bf x}-{\bf y}|})=-2\int_0^{\infty}
\frac{d\bf p}{\bf p} \left[\frac{\frac{1}{\pi}\hat{V}({\bf
p})+1}{2\sqrt{\frac{2}{\pi}\hat{V}({\bf
p})+1}}-\frac{1}{2}\right]\nonumber\\ \times [1-\cos(|{\bf x}-{\bf
y}|.{\bf p})] \eeqn

\noindent where we have used $p_{1}={\bf p}$. It is convenient to
check the validity of this equation by considering a non trivial
potential such as the Coulomb interaction
\begin{equation}
V(|{\bf x}-{\bf y}|)=\frac{e^2}{\sqrt{|{\bf x}-{\bf y}|^2+d^2}}\
\end{equation}

\noindent with Fourier transform given by

\begin{equation}
\hat{V}({\bf p})=2e^2K_0(d{\bf p}).
\end{equation}

\noindent where $d$ is a very small length scale. This computation
is interesting for two reasons. Firstly, one should stress that
equation (\ref{I_{R,L}}) has been checked only for local
(short-range) potentials: the covariant Thirring model (see
eq.(\ref{eq:Thirring})) and the short range TL model \cite{non
local Thirring}. On the other hand, the electronic correlator for
a Coulomb potential has been computed by Schulz \cite{Schulz2}
using standard (operational) bosonization \cite{conventional
bosonization} and thus our computation will give an independent
confirmation by means of a different approach. To achieve this
goal we consider a large distance approximation of (\ref{I_{R,L}})
that consists in inserting $|{\bf x}-{\bf y}|^{-1}$ in the lower
limit of (\ref{eq:integral}) and disregarding the cosine term,
which is valid for $\frac{1}{d}>>\frac{1}{|{\bf x}-{\bf y}|}$, we
obtain:

\begin{equation}\label{eq:ShulzFormula}
G_{R,L}({\bf x}-{\bf y})=\pm i C(\Lambda)sign({\bf x}-{\bf y})
\exp{\left[-\alpha\left(\log\frac{|{\bf x}-{\bf y}|}{\beta
d}\right)^{3/2}\right]}
\end{equation}

\noindent where $\alpha$ and $\beta$ are two constants that depend
on $e^2$ and $C(\Lambda)$ depends on an ultraviolet cutoff
$\Lambda$. This is our first non trivial contribution. We have
obtained, as anticipated, the behavior previously found by Schulz
\cite{Schulz2}. In passing we note that a similar behavior has
been previously observed by considering a Coulomb interaction
between the core-hole and the conduction electrons
\cite{Chen-Lee}.

Now we undertake the computation of $G_{R,L}({\bf x}-{\bf y})$ for
the so called "super long-range" potential given by
(\ref{eq:SuperLR}), whose Fourier transform is
\begin{equation}
\hat{V}({\bf
p})=2e^2\cos\left(\frac{\pi}{2}\epsilon\right)\Gamma(\epsilon)(d|{\bf
p}|)^{-\epsilon}.
\end{equation}

\noindent Inserting this expression in (\ref{eq:integral}) one
gets
\begin{equation}
I(\xi)=-\int_{\lambda}^\infty\frac{d|{\bf q}|}{|{\bf
q}|}\left[\frac{|{\bf q}|^{-\epsilon}+2}{2\sqrt{|{\bf
q}|^{-\epsilon}+1}} -2\right] [1-\cos (\xi |{\bf q}|)]
\end{equation}

\noindent where it has been natural to define the new variable
\begin{equation}\label{eq:Xi}
\xi=\left[\frac{4}{\pi} e^2\cos\left(\frac{\pi}{2}\epsilon\right)
\Gamma(\epsilon)\right]^{1/\epsilon}\frac{{\bf x}-{\bf y}}{d}.
\end{equation}

\noindent Note that we have also introduced a small quantity
$\lambda$ to be set equal to zero at the end of our computation.
Although $I(\xi)$ is infrared convergent, it is convenient to
split it into two pieces: $I(\xi)=I_1(\xi)-I_1(\xi=0)$ with
\begin{equation}
I_1(\xi)=\frac{1}{2}\int_\lambda^\infty\frac{d|{\bf q}|}{|{\bf
q}|} \left[\frac{|{\bf q}|^{-\epsilon}+2}{\sqrt{|{\bf
q}|^{-\epsilon}+1}} -2\right]\cos(\xi |{\bf q}|).
\end{equation}

\noindent The first $\xi$-independent piece can be very easily
computed yielding
\begin{equation}
I_1(\xi=0)=-\frac{1}{\epsilon}\lambda^{-\epsilon/2}-\log
\lambda-\frac{1}{\epsilon}(\log4-1) + O(\lambda^{\epsilon/2}).
\end{equation}

\noindent The evaluation of $I_1(\xi)$ is more involved, as
expected. It can be achieved by splitting the integration region
into two sub-intervals ($\lambda,1$) and ($1,\infty$), performing
then series expansions of the corresponding square roots factors
and finally integrating term by term. Putting all this together
and taking the limit ${\lambda}\rightarrow0$ one finally obtains
\beqn\label{eq:exact} I(\xi)&=\frac{1}{4}\sum_{n=0}^\infty(2a_n +
a_{n+1})
\frac{1}{\mu_n}\left[\Phi(\mu_n,\mu_n+1;i|\xi|)+\Phi(\mu_n,\mu_n+1;-i|\xi|)\right]\nonumber\\
&+\frac{1}{4}\sum_{n=0}^\infty(a_n + 2a_{n+1})|\xi|^{\nu_n}\left[
e^{i\frac{\pi}{2}\nu_n}\Gamma(-\nu_n,i|\xi|)+
e^{-i\frac{\pi}{2}\nu_n}\Gamma(-\nu_n,-i|\xi|)\right]\nonumber\\
&-\frac{1}{4}|\xi|^{\epsilon/2}\left[e^{-i\frac{\pi}{4}\epsilon}
\Gamma(-\frac{\epsilon}{2},-i|\xi|)+e^{i\frac{\pi}{4}\epsilon}
\Gamma(-\frac{\epsilon}{2},i|\xi|)\right] - Ci(|\xi|)\nonumber\\
&+\frac{1}{2}|\xi|^{\epsilon/2}\Gamma(-\frac{\epsilon}{2})\cos(\frac{\pi}{2}\epsilon)
+\log|\xi|+\left(C+\frac{1}{\epsilon}-\frac{1}{\epsilon}\log4\right)
\eeqn

\noindent where $\Phi(a, b, z)$ is Kummer's confluent
hypergeometric function, $\Gamma(\alpha,x)$ is the incomplete
Gamma function, and $Ci(z)$ is the cosine integral function. We
have also defined the coefficients
$\mu_n=(n+\frac{1}{2})\epsilon$, $\nu_n=(n+1)\epsilon$, and
$a_n=\frac{(-1)^n\Gamma(n+\frac{1}{2})}{\Gamma(\frac{1}{2})\Gamma(n+1)}$.
Formula (\ref{eq:exact}) is our main result. Indeed, by recalling
that
\begin{equation}
G_{R,L}({\xi})=G_{R,L}^0({\xi})\,\ \exp{I(\xi)},
\end{equation}
\noindent one sees that (\ref{eq:exact}) gives a complicated but
analytical and exact result for the fermionic propagator of the TL
model in the presence of the long-range interaction given by
equation (\ref{eq:SuperLR}).

Of course, it is now interesting to study the long-distance
behavior of our solution ($|\xi|\gg1$). A careful evaluation of
the dominant contributions to eq.(\ref{eq:exact}) for this case
yields
\begin{equation}\label{eq:long distance}
G_{R,L}(\xi)\propto\pm i \,\ sign(\xi)\exp\left[
\frac{1}{2}\Gamma(-\frac{\epsilon}{2})\cos(\frac{\pi}{4}\epsilon)|\xi|^{\epsilon/2}\right]
\end{equation}
\noindent Note that, exactly as it happens with Schulz's solution,
the above function decays faster that any power-law. Moreover, the
larger the range of the potential (larger $\epsilon$) the faster
is the Green function decay. Let us also mention that the
definition of the long-distance regime is different in both cases
(see eq.(\ref{eq:Xi})), since the quantity $\epsilon$ determines
the range of the potential. This does not prevent us from
comparing both long-distance decays for definite values of
$\epsilon$. A simple numerical comparison allows to verify that
the results are increasingly similar when distances increase.

It is also interesting to notice that the long-distance regime can
be obtained following a much simpler route, analogous to the one
employed to get this regime for the Coulomb case (see the
paragraph preceding eq.(\ref{eq:ShulzFormula})). However, we must
mention that the condition that allows to drop the oscillating
integrals leads, in the present case, to a result that is only
valid for small $\epsilon$. Under this condition, replacing
(\ref{eq:SuperLR}) in (\ref{eq:integral}) and disregarding the
cosine term gives
\begin{equation}
G_{R,L}(\xi)\propto \pm i \,\ sign(\xi)
\exp\left[-\frac{1}{\epsilon}|\xi|^{\epsilon/2}\right],
\end{equation}

\noindent which coincides with (\ref{eq:long distance}) for
$\epsilon\ll1$. We want to stress here that this agreement is an
important consistency check of our exact result since it has been
derived in a very straightforward way, {\it without using equation
(\ref{eq:exact})}.

\vspace{1cm}

In summary, we have shown how to compute, through path-integral
methods, the equal-time fermionic one-particle Green function in a
simple version of the TL model. In particular we have obtained a
rather involved but exact analytical expression for this
propagator in the case of a  "super long-range" potential (see
(\ref{eq:SuperLR})). This result is indeed of academic interest,
since most of the explicit results found in the literature
actually correspond to long distance regimes (see for instance
\cite{Schulz2}). In passing, and as a consistency check of our
computation, we have given a path-integral derivation of the Green
function corresponding to the Coulomb case, a well-known result
previously found by Schulz.

\newpage {\bf
Acknowledgements}

A.I. is partially supported by a Fellowship for students of
Comisi\'on de Investigaciones Cient\'{\i}ficas de la Provincia de
Buenos Aires (CICPBA), Argentina.

C.N. is partially supported by Universidad Nacional de La Plata
(UNLP) and Consejo Nacional de Investigaciones Cient\'{\i}ficas y
T\'ecnicas (CONICET), Argentina. He thanks Eduardo Fradkin for
calling his attention to ref.\cite{Schulz2}.

Both authors are grateful to Yang Chen for calling their attention
to ref. \cite{Chen-Lee}.


\begin{thebibliography}{99}
\bibitem{organic conductors} D. Jerome, H. Schulz, Adv. Phys.
{\bf31}, 299 (1982).
\bibitem{salts} A.J. Epstein et al., Phys. Rev. Lett. {\bf47}, 741
(1981).
\bibitem{quantum wires} A.R. Goni et al., Phys. Rev. Lett. {\bf67},
3298 (1991).
\bibitem{Hubbard} See, for instance, E. Fradkin, Field Theories of
Condensed Matter Systems, Addison-Wesley Publishing Co., 1991.
\bibitem{Solyom} J. Solyom,  Adv. Phys. {\bf 28}, 209 (1979).
\bibitem{Haldane} F.D.M. Haldane, J. Phys. C {\bf 14}, 2585, (1981).
\bibitem{Schulz1} H.J. Schulz, J. Phys. C {\bf 16}, 6769, (1983).
\bibitem{quasi-exp} B. Dardel et al., Europhys. Lett. {\bf 24},
687 (1993); M.Nakamura et al., Phys. Rev. B {\bf 49}, 16191(1994);
A. Sekiyama et al., ibid. {\bf 51}, 13899 (1995).
\bibitem{Fermi-edge} H.Otani and T. Ogawa, Phys. Rev. B {\bf 54},
4540 (1996).
\bibitem{insulator-metal} D. Poilblanc, S. Yunoki, S. Maekawa, E.
Dagotto, Phys. Rev. B {\bf 56}, R1645 (1997).
\bibitem{lattice} G.Fano, F. Ortolani, A. Parola, L. Ziosi,
cond-mat/9909140, to be published in Phys. Rev. B.\\ S. Capponi,
D. Poilblanc, T. Giamarchi, cond-mat/9909360.
\bibitem{Schulz2} H.J.Schulz, Phys.Rev.Lett.{\bf 71}, 1864, (1993).
\bibitem{conventional bosonization} V.J.Emery, in Highly
Conducting One-dimensional Solids, edited by J.T. Devreese et al.
(Plenum, New York, 1979), p. 327; M. Stone, Bosonization (World
Scientific, Singapore, 1994); R. Shankar, in Low Dimensional
Quantum Field Theories for Condensed Matter Physicists, edited by
Lu Yu, S. Lundqvist and G. Morandi (World Scientific, Singapore,
1995).
\bibitem{path-integral bosonization} R.Gamboa-Sarav\'{\i}, F.Schaposnik and J.Solomin,
Nucl. Phys. B {\bf185}, 239 (1981).\\ K.Furuya,
R.Gamboa-Sarav\'{\i} and F.Schaposnik, Nucl. Phys.B {\bf208}, 159
(1982).\\ C.Na\'on, Phys. Rev. D {\bf31}, 2035 (1985).
\bibitem{non local Thirring}C.Na\'on, M.C.von.Reichenbach and M.L. Trobo,
             Nucl. Phys. {\bf B 435} [fs] 567 (1995);ibid. {\bf B 485} [fs] 665
             (1997).\\
M.Man\'{\i}as, C.Na\'on , M.L.Trobo, Nucl. Phys. B 525 [FS], {\bf
721} (1998).
\bibitem{Schmeltzer} D. Schmeltzer, Phys. Rev. B {\bf 52}, 7939 (1995).

\bibitem{Ground} D. G. Barci, C. M. Na\'on, Int. J. Mod. Phys. A {\bf 13}, 1169
(1998).
\bibitem{Chen-Lee} Y. Chen and D. Lee, Phys. Rev. Lett. {\bf 69},
1399 (1992).
\bibitem{Gradshteyn} I. S. Gradshteyn y I. M. Ryzhik, {\em Table of
Integrals, Series and Products}, Academic Press, Fifth edition
(1994).
\end{thebibliography}
\end{document}